\documentclass[conference]{IEEEtran}

\IEEEoverridecommandlockouts
\usepackage{caption}

\usepackage{threeparttable}
\usepackage{tablefootnote}

\usepackage{algorithm}
\usepackage[noend]{algpseudocode}

\usepackage{varwidth}
\usepackage{amsmath}
\usepackage{color} 
\usepackage{hyperref}

\usepackage{graphicx,epstopdf,algpseudocode,caption,url}   
\usepackage{multirow}  
\graphicspath{{./images}}

\usepackage{arydshln} 

\begin{document}

\title{Federated Learning Attacks and Defenses: A Survey}

\author{Yao Chen$ ^{1} $, Yijie Gui$ ^{1}$, Hong Lin$ ^{1}$, Wensheng Gan$ ^{1,2*}$\thanks{\IEEEauthorrefmark{1}Corresponding author.}, Yongdong Wu$ ^{1}$ \\ 
	\\
	$ ^{1} $Jinan University, Guangzhou 510632, China\\
	$ ^{2} $Pazhou Lab, Guangzhou 510330, China\\

	Email: \{csyaochen, y.j.gui123, lhed9eh0g, wsgan001, wuyd175\}@gmail.com
}

\maketitle

\begin{abstract}

In terms of artificial intelligence, there are several security and privacy deficiencies in the traditional centralized training methods of machine learning models by a server. To address this limitation, federated learning (FL) has been proposed and is known for breaking down ``data silos" and protecting the privacy of users. However, FL has not yet gained popularity in the industry, mainly due to its security, privacy, and high cost of communication. For the purpose of advancing the research in this field, building a robust FL system, and realizing the wide application of FL, this paper sorts out the possible attacks and corresponding defenses of the current FL system systematically. Firstly, this paper briefly introduces the basic workflow of FL and related knowledge of attacks and defenses. It reviews a great deal of research about privacy theft and malicious attacks that have been studied in recent years. Most importantly, in view of the current three classification criteria, namely the three stages of machine learning, the three different roles in federated learning, and the CIA (Confidentiality, Integrity, and Availability) guidelines on privacy protection, we divide attack approaches into two categories according to the training stage and the prediction stage in machine learning. Furthermore, we also identify the CIA property violated for each attack method and potential attack role. Various defense mechanisms are then analyzed separately from the level of privacy and security. Finally, we summarize the possible challenges in the application of FL from the aspect of attacks and defenses and discuss the future development direction of FL systems. In this way, the designed FL system has the ability to resist different attacks and is more secure and stable.

\end{abstract}


\IEEEpeerreviewmaketitle

\section{Introduction}  \label{sec:introduction}

With the vigorous development of big data and artificial intelligence, large amounts of data and models have been generated. The process and the transfer of data have become much more frequent in the meantime. As we all know, in many industries, data often exists in the form of silos, and the most straightforward way to solve the silo problem is to ensemble data into one party for further processing \cite{yang2019federated}. However, this practice will certainly cause data privacy leakage problems. Nowadays, many countries have made efforts to strengthen the protection of citizens' private data security. Take the European Union's General Data Protection Regulation (GDPR) as an example, which came into force on May 25, 2018. Protecting users' personal privacy and data security is the goal of this regulation. Similarly, since 2017, the Cybersecurity Law of the People's Republic of China has guaranteed cybersecurity, protected the legitimate rights and interests of citizens, and promoted the healthy development of economic and social information technology\footnote{\url{http://www.gov.cn/xinwen/2016-11/07/content_5129723.htm}}. Naturally, with the emphasis on data privacy and security becoming a worldwide trend, breaking down data silos and making full use of data have become hot topics today. To address this limit, federated learning (FL) \cite{zhang2021survey,li2020federated,kairouz2021advances} is proposed as a data integration method that complies with data privacy and security laws.

Google proposed FL in 2016 \cite{mcmahan2017communication} to address the limitation of updating models locally by the user using an Android mobile phone. This technique has been used in many other areas in combination with other expertise, such as enterprise data alliance \cite{yang2020prototyping}, blockchain \cite{lu2019blockchain}, smart finance \cite{long2020federated}, smart healthcare \cite{kumar2021federated}, etc. FL is actually an encrypted distributed machine learning technique that can effectively assist multiple organizations and the usage of data while meeting the requirements of privacy protection, data security, and government regulations. It is characterized by the following four parts: multi-party collaboration; equality of all parties; data privacy protection; and data encryption. A concept that is similar to FL is joint learning \cite{finkel2010hierarchical}. The difference between them is that joint learning has no requirements for aggregated methods or privacy about data. In addition, there is still a difference between FL and privacy protection theories commonly used in data mining \cite{gan2017data,gan2021survey}, such as differential privacy \cite{dwork2008differential,abadi2016deep,li2022frequent}, k-anonymity \cite{sweeney2002k,lefevre2006mondrian}, and l-diversity \cite{machanavajjhala2007diversity}. Specifically, FL protects users' data privacy through the exchange of parameters under specific encryption mechanisms. However, differential privacy, k-anonymity, and l-diversity methods protect the users' privacy information by adding noise to data or obfuscating sensitive attributes in the databases.

\begin{figure*}[h]
    \centering
    \includegraphics[scale=0.434]{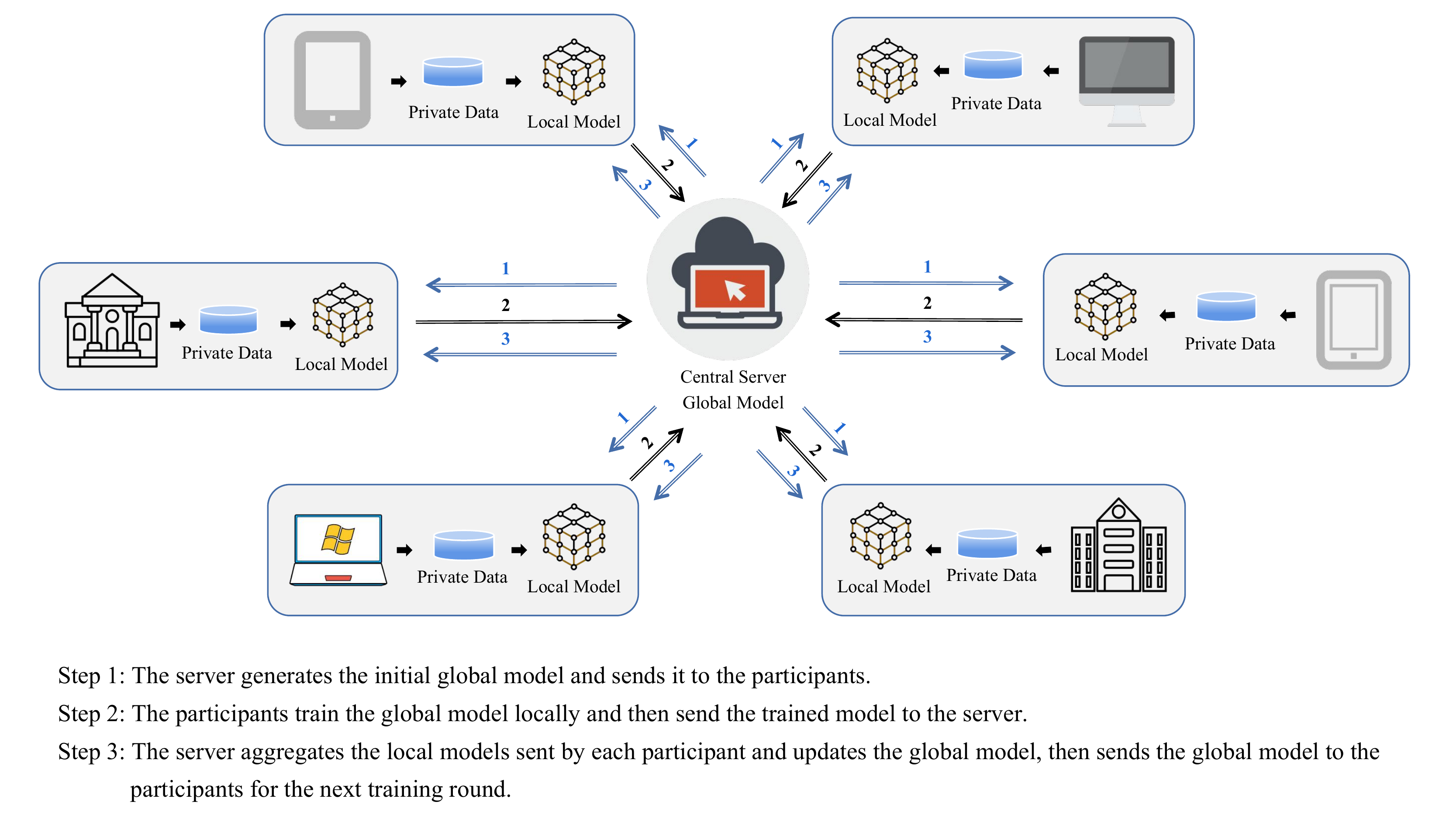}
    \caption{An overall flow of FL \cite{mothukuri2021survey}.}
    \label{fig:flow}
\end{figure*}

The workflow of FL can be generally divided into three steps: the server generates the initial global model, the participants train the model locally, and the server updates the global model. This is shown in Fig. \ref{fig:flow}. On the basis of the distribution of data sources among all participants, FL can be divided into three categories \cite{yang2019federated}: horizontally federated learning (HFL), vertically federated learning (VFL), and federated transfer learning (FTL). The essence of HFL is the union of samples. When users' features of two datasets overlap more and users overlap less, we divide the dataset horizontally and take out the part of the data where the users' features of both datasets are the same, but the users are not exactly the same for training (that is the user dimension). The nature of VFL is the union of features. When two datasets have more overlap in users but less overlap in users' features, we divide the dataset vertically and take out the part of the data where the users on both sides are the same, but the user features are not exactly the same for training (that is the feature dimension). However, in those cases where both datasets have less overlap in users and users' features, we do not slice the data and use the transfer learning method to overcome the shortfall of data or label. The transfer learning method can be used by alliances between banks and e-commerce companies in different regions. FTL is applicable to some scenarios based on deep neural networks.

Although FL does not directly exchange data and has a higher privacy guarantee than traditional machine learning, there are still some insecure problems in specific practical applications that need to be further studied and solved \cite{liu2020secure}. It faces four challenges in general: high communication costs, systemic heterogeneity, statistical heterogeneity, and data security \cite{li2020federated}. The first three challenges are problems that FL may encounter in practical applications. The last challenge ensures that federal learning meets privacy, security, and various legal regulations. In naive FL, the confidentiality of data mainly depends on the use of cryptographic algorithms, such as AES \cite{daemen1999aes}, SM4 \cite{zhang2016security}, RSA \cite{wiener1990cryptanalysis}, etc. These algorithms enable the plaintext data to be encrypted to obtain irreversible ciphertext data, ensuring that the privacy of the data is not leaked. However, there are some research works \cite{ganju2018property,ateniese2015hacking,melis2019exploiting,shokri2017membership} proving that some private data can be inferred from the transmission of data. The member inference attack was first proposed by Shokri \textit{et al.} \cite{shokri2017membership} and aims to use the trained model to determine whether a sample belongs to the corresponding training set. This can reveal private information in certain situations, such as disease classification models in the medical field. As attackers continue to find out vulnerabilities in the network, the model inversion attack \cite{fredrikson2015model}, the adversarial attack \cite{goodfellow2014explaining}, the backdoor attack \cite{li2022backdoor,wang2020attack}, the Denial of Service attack \cite{garber2000denial}, and other attacks are successively discovered by them. In order to cope with various attacks, differential privacy \cite{dwork2008differential,dwork2014algorithmic}, secure multi-party computing \cite{goldreich1998secure,zhao2019secure}, homomorphic encryption \cite{naehrig2011can,martins2017survey}, and other strategies for defense have also been proposed to ensure security in FL.

Until now, most of the existing literature reviews about FL attacks and defenses mainly focused on specific attack types \cite{bagdasaryan2020backdoor, luo2021feature} and listed each attack type for analysis and explanation \cite{jere2020taxonomy,mothukuri2021survey,lyu2020threats}. Some more specific literature classifies attacks according to different criteria. For example, Liu \textit{et al.}  \cite{liu2022threats} divided threat models, attacks, and defenses through three stages of FL (data and behavior auditing phase, training phase, and prediction phase). Chen \textit{et al.}  \cite{Chen2022threats} classified attacks according to three different principles of information security (i.e., confidentiality, integrity, and availability). Wu \textit{et al.} \cite{jianhan2022threats} enumerated types of attacks according to the potential attackers in the FL system. One of the highlights of this paper is to look for the intersection of the above three criteria and combine them reasonably.

\textbf{Contributions}: This paper aims to summarizing the development of attack and defense technologies in the frameworks of different FL systems in recent years. The classification of existing attacks, which combines the current three classification criteria, is highlighted in this paper. In addition, we categorize and discuss various attack strategies in detail, including poisoning attack, inference attack, reconstruction attack, etc. Specifically, existing defenses are discussed in detail from the perspective of privacy and security, respectively. Finally, we put forward six research directions for the future development of attack and defense in various FL systems.

\textbf{Outline}: The remaining content of this paper is organized and presented as follows: we provide and describe some fundamental knowledge about FL in Section \ref{sec:background}. Most importantly, we categorize attacks and defenses and give a detailed analysis in Section \ref{sec:attacks} and Section \ref{sec:defenses}. Furthermore, in Section \ref{sec:challenges}, we present a comprehensive review of challenges and potential future research directions for FL. Finally, we make a conclusion for this review in Section \ref{sec:conclusion}.

\section{Background}  \label{sec:background}

Federated learning is a machine learning paradigm that involves multiple participants \cite{liu2022threats}. In this paradigm, the participants can build the model jointly without disclosing the underlying data or its cryptographic form. In short, FL implements training and builds a shared model without letting data leave the local area. This section provides background knowledge on FL attacks and defenses, including different types of FL and some threats to security and privacy that FL encounters in each process.

\subsection{Types of Federated Learning}

Based on the data characteristics of the participants and the distribution of the data samples, FL can be classified into the following three types \cite{yang2019federated}.

\textbf{Horizontally federated learning (HFL).} The characteristic of HFL is that participants have different data samples, but the data features in the samples overlap \cite{liu2022threats}. For example, given two hospitals in different locations, the intersection of the data samples from these two hospitals is small. However, the business of hospitals is very similar. Therefore, there are many overlaps in the data characteristics of the data samples. In this case, we can use HFL to build a joint model.

\textbf{Vertically federated learning (VFL).} The characteristic of VFL is that data samples owned by participants overlap, but the data features in the samples are different \cite{qammar2022federated}. For instance, a bank and an e-commerce company are in the same area. Their user groups include some local residents, so the intersection of the data samples from these two institutions is large. The characteristics of the data collected by the bank are related to the credit level of the user. And the characteristics of the data collected by the e-commerce company are related to consumer behaviors. Therefore, the characteristics of the data collected by these two institutions are not the same. This situation is suitable for the VFL.

\textbf{Federated transfer learning (FTL).} FTL is characterized by less overlap of data characteristics in different data samples and less overlap of data samples owned by the participants \cite{liu2020secure}. FTL combines the ideas of FL and transfer learning, where transfer learning is a machine learning method that transfers knowledge learned in one domain or task to another domain or task \cite{pan2009survey, jing2019quantifying}. FTL uses transfer learning to overcome the problem of non-overlapping samples and non-overlapping features \cite{liu2017understanding}.

\subsection{Security and Privacy Threats in FL}

\begin{figure*}[h]
    \centering
    \includegraphics[scale=0.43]{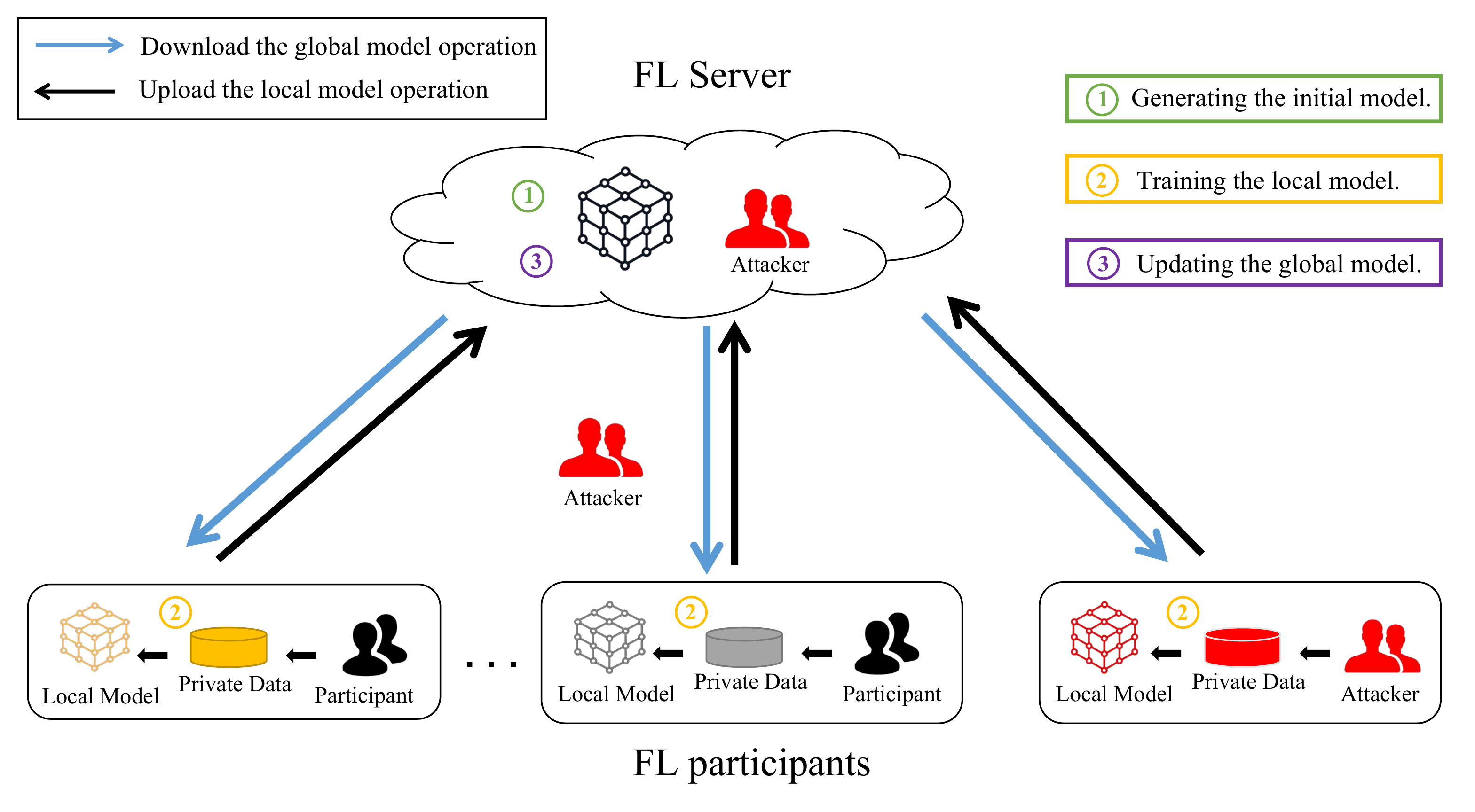}
    \caption{Threats in a federated learning environment.}
    \label{fig:FLprocess}
\end{figure*}

Fig. \ref{fig:FLprocess} depicts the classic workflow of FL training models and the privacy and security threats that exist.
The workflow of FL training models is generally divided into three steps: generate the initial model, train the local model, and update the global model.

\textbf{Generating the initial model.} FL usually starts with an initial global model generated by the server. This global model is then broadcast to those participants who trained the model together \cite{qammar2022federated}. In this step, a malicious server may inject malicious data into the global model. This behavior affects the training and prediction of the model. All participants download the global model from the malicious server and upload the local model to the malicious server. This attack behavior may threaten the privacy and security of the participants by attacking the models uploaded by the participants. Poisoning attacks often occur during this step.

\textbf{Training the local model.} Each participant downloads a global model from the server. Then, the participant trains this global model with local data. This training process is done locally. As a result, raw data from participants does not leave the local area. Finally, each participant uploads the trained local model to the server. In this step, the attacker may be a participant who injects malicious behavior into the local model or an eavesdropper who steals information from a communication channel. In other words, the attacker, as a participant, influences the training of the FL by sending local models with malicious behaviors to the server \cite{bagdasaryan2020backdoor}. Besides, this attacker can also infer some sensitive information from the global model that the server sends each time. The attacker can also use the information stolen from the communication channel to carry out the attack. The communication channel transmits the local models uploaded by the participants and the global model broadcast by the server. During this uploading and downloading process, an attacker may tamper with or steal the models, thus affecting the training of the global model and compromising the privacy of the participants. Poisoning attacks and inference attacks are common here.
    
\textbf{Updating the global model.} The server collects the local models uploaded by all participants and then performs a model aggregation operation. This step's goal is to recombine all of the collected local models into a global model. The aggregation algorithm is involved here. FedAvg is the classical aggregation algorithm \cite{mcmahan2017communication}, which obtains a new common model by weighting the parameters of all the models and averaging them. In this step, the server may be injected with malicious behavior by the attacker, which can affect the training of the model \cite{lyu2020threats}. In addition to this, the malicious server receives local models from all participants. The attacker can use the local models uploaded by the participants to perform inference attacks, or he can use the generated models to reconstruct the training set of the participants. These actions may infer sensitive information about the participants and thus threaten the privacy and security of the participants.

\section{Attacks in FL}  \label{sec:attacks}

There are many types of attacks in FL, with different standards for the classification of attacks in previous studies. For instance, Chen \textit{et al.} \cite{Chen2022threats} classified attacks based on three different principles of information security named CIA (confidentiality, integrity, and availability). Liu \textit{et al.} \cite{liu2022threats} listed the corresponding possible attacks based on the three phases of the FL (including data auditing, training, and predicting). Wu \textit{et al.} \cite{jianhan2022threats} enumerated types of attacks according to the potential attackers of the FL system (including local workers, the central server, and the eavesdropper). In this section, we categorize and introduce the major attacks using the above criteria.  As shown in Table \ref{fig:Classification}, we retain two classification standards: potential attack roles and three principles of information security. Since most attacks exist in the latter two phases, we remove the data auditing phase \cite{liu2022threats}. In addition, FL is usually in the training phase in most cases. However, we believe that for attackers, when the model is trained to a certain extent, they can enter the prediction stage and carry out some attacks through this stage, even though the model is not fully trained. Therefore, in our opinion, the prediction stage is also an important part of federated learning.

\begin{table}[h]
    \centering
    \caption{Classification of attacks in FL}
    \includegraphics[trim=0 0 0 0,clip,scale=0.31]{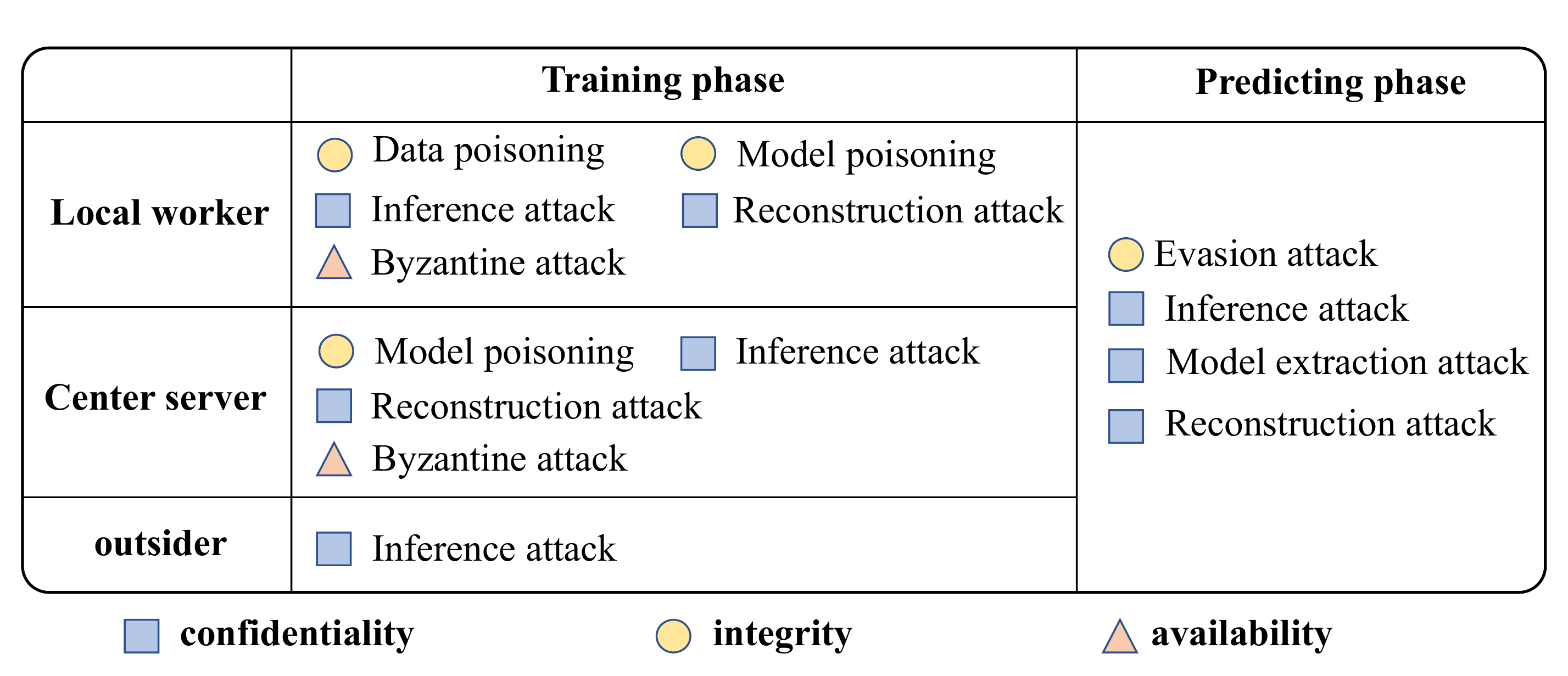}
    \label{fig:Classification}
\end{table}

\subsection{Poisoning Attacks}

There are some different classification standards for poisoning attacks. Specifically, they can be divided into data poisoning and model poisoning according to the poisoning objects, and can be divided into random attacks and backdoor attacks (target attacks) according to whether the attack is targeted. It violates the integrity principle of FL by making the global model unusable under certain conditions.

Data poisoning refers to an attacker tampering with or adding data to the training set maliciously, which eventually leads to the destruction or hijacking of the model. This attack is usually implemented by the local worker, who is the owner of the data. Later, the idea of clean-label poisoning comes up \cite{zhao2020clean}. This is a backdoor attack that adds malicious data instead of changing it. Conversely, dirty data poisoning usually involves changing the original dataset through label-flipping \cite{fung2018mitigating, tolpegin2020data} or the generation of toxic samples \cite{chen2017targeted, gu2019badnets}.

Model poisoning means that the attacker changes the parameters of the target model directly, causing errors in the global model or leaving a backdoor. It has been shown to be more effective than data poisoning \cite{bhagoji2019analyzing}. Perturbing the weights of convolutional neural networks in a targeted manner can be used to insert backdoors \cite{dumford2020backdooring}. By using available bit-flipping techniques, the target model can be converted into the Trojan infection model \cite{rakin2020tbt}. Since the model needs to be transmitted repeatedly between the local worker and the center server, either one of them could be an attacker.

\subsection{Inference Attacks}

Although the data transmitted from the client to the server is not the original data, there is still a risk of leakage. Inference attacks indicate that attackers can use the eavesdropped information to infer useful information to a certain extent, which obviously destroys the confidentiality of the model. The member inference attack was first proposed in \cite{shokri2017membership}, which trains a model with similar functions on a specified sample and then judges whether the sample is used for the target model's training. This training method is called the Shadow training technique. What's more, the attribute inference attack \cite{ateniese2015hacking} aims to determine whether a participant's data is relevant to a certain attribute. There have been studies to prove that this attack can infer the accent information of the training set in FL of speech recognition. According to a recent study \cite{wainakh2021user}, sensitive information represented by labels can potentially be inferred from user-uploaded gradients.

\subsection{Reconstruction Attacks}

Unlike inference attacks that cannot reveal raw data, reconstruction attacks can obtain the original information of the training dataset by collecting some information such as predicted confidence values, model parameters, and gradients. Therefore, this type of attack is also a confidentiality attack. The model inversion attack is one of the reconstruction attacks and was first proposed by \cite{fredrikson2014privacy}. This model shows that when the adversary has white-box access rights, the adversary can use the linear model to estimate the attribute information of the original data. Another attack method is generative adversarial networks (GAN) \cite{hitaj2017deep} and it points out that attackers can obtain samples of other participants, and this process only requires black-box access. And Deep Leakage from Gradients (DLG) shows that the attacker could recover the original data by analyzing the gradient information \cite{zhu2019deep}. Then, the Improved Deep Leakage from Gradients (IDLG) \cite{zhao2020idlg} improves the accuracy of data recovery. Inverting gradients based on DLG were proposed \cite{geiping2020inverting} and it broadened the attack scenario to include the actual industrialized scenario rather than being limited to the strong assumption of low-resolution recovery and a shallow network. Recently, Generative Regression Neural Network (GRNN) based on GAN was proposed to restore the original training data without the need for additional information \cite{ren2022grnn}. It indicated that GRNN has stronger robustness and higher accuracy than previous methods.

\subsection{Model Extraction Attack}

When FL has finished training the model, the global model will serve outsiders in the form of an API. At this time, the user may query the relevant information of the target model through the API loop and finally achieve the effect of extracting the model. Tramèr \textit{et al.} \cite{tramer2016stealing} first demonstrated that this attack will be effective when the attacker has the same distribution of data and model-related information as the model. The attack proposed in \cite{wang2018stealing} can obtain hyperparameters located at the bottom layer of the model. Orekondy \textit{et al.} \cite{orekondy2019knockoff} proposed the Knockoff Net, through which attackers can steal based on the confidence value output by the API, and the stealing effect is positively correlated with the complexity of the target model.

\subsection{Evasion Attacks}

Evasion attack is a type of attack in which an attacker deceives the target machine learning system by constructing specific input samples without changing the target machine learning system. It usually occurs during the prediction phase, when the model has finished training. The effect of this kind of attack can be summarized as the model extrapolates the original answer ``A" to be the wrong answer ``B". Main feature of it is a wide spread of hazards, including road sign recognition for autonomous driving \cite{lu2017adversarial}, face recognition \cite{sharif2016accessorize}, voice recognition system \cite{carlini2016hidden}, etc. Evasion attack is an integrity attack due to the spoofing of the model.

\subsection{Byzantine Attacks}

Byzantine attack is a type of attack in which an attacker hides among the participants of FL and arbitrarily uploads malicious data, aiming to destroy the global model (e.g., model availability). To deal with this attack, it is common to combine stochastic gradient descent (SGD) with different robust aggregation rules (e.g., Krum, Median). However, the stochastic gradient noise induced by SGD makes it impossible for the server to judge whether it is malicious information or the noise of real information, which becomes an exploit point for attackers \cite{wu2020federated}. Xie \textit{et al.} \cite{xie2020fall} use a method called ``inner product manipulation" to make the aggregated vector direction in the server inconsistent with the true gradient, thereby causing SGD to fail. Similar to this idea, it has been proven that by poisoning the local model, the global model has a large test error rate \cite{fang2020local}.

\textbf{Discussion:} Whether an attack is significant depends on the depth of its damage and the scope of its damage. We can see that each of the above attacks makes assumptions about the attacker's privileges or attack capabilities. Obviously, an excellent attack should have as few assumptions as possible, which mean that it is more applicable to a wider range of scenarios and more difficult to be detected by defense strategies. Therefore, making fewer assumptions should be a principle for designing future attack strategies, and the corresponding defense strategy should also detect the difference between malicious participants and attacking participants as much as possible. In conclusion, comprehensive and in-depth attack research can promote the development of defense.

\section{Defenses in FL} \label{sec:defenses}

In this section, we describe several defense methods in the FL environment at the privacy level and the security level, respectively.
The goal of the security-based defense policy is different from that of the privacy-based defense policy. More specifically, privacy refers to private information that a person does not want others to know and invade, focusing on sensitive personal information \cite{lyu2020privacy}; security focuses on confidential data and information assets, not just personal information. As we all know, the CIA follows three core security principles \cite{alazab2021federated}. The attack on FL
is considered at the privacy level, when the attacker tries to infer private information about the participant. Privacy protection methods are used to defend against privacy attacks and to ensure that sensitive data is not leaked to others \cite{yin2021comprehensive}. The security attack is a malicious action performed by hackers with specialized knowledge to compromise the confidentiality, integrity, and availability of data and models. And the security defense aims to improve the FL framework's CIA.

\subsection{Defenses at the Privacy Level}

In this section, some privacy protection methods are discussed for defending against attacks at the privacy level. Table \ref{fig:Defend} describes these approaches, including the types of attacks they defend against and their shortcomings.

\begin{table}[h]
    \centering
    \caption{Privacy protection methods in FL.}
    \includegraphics[trim=0 0 0 0,clip,scale=0.41]{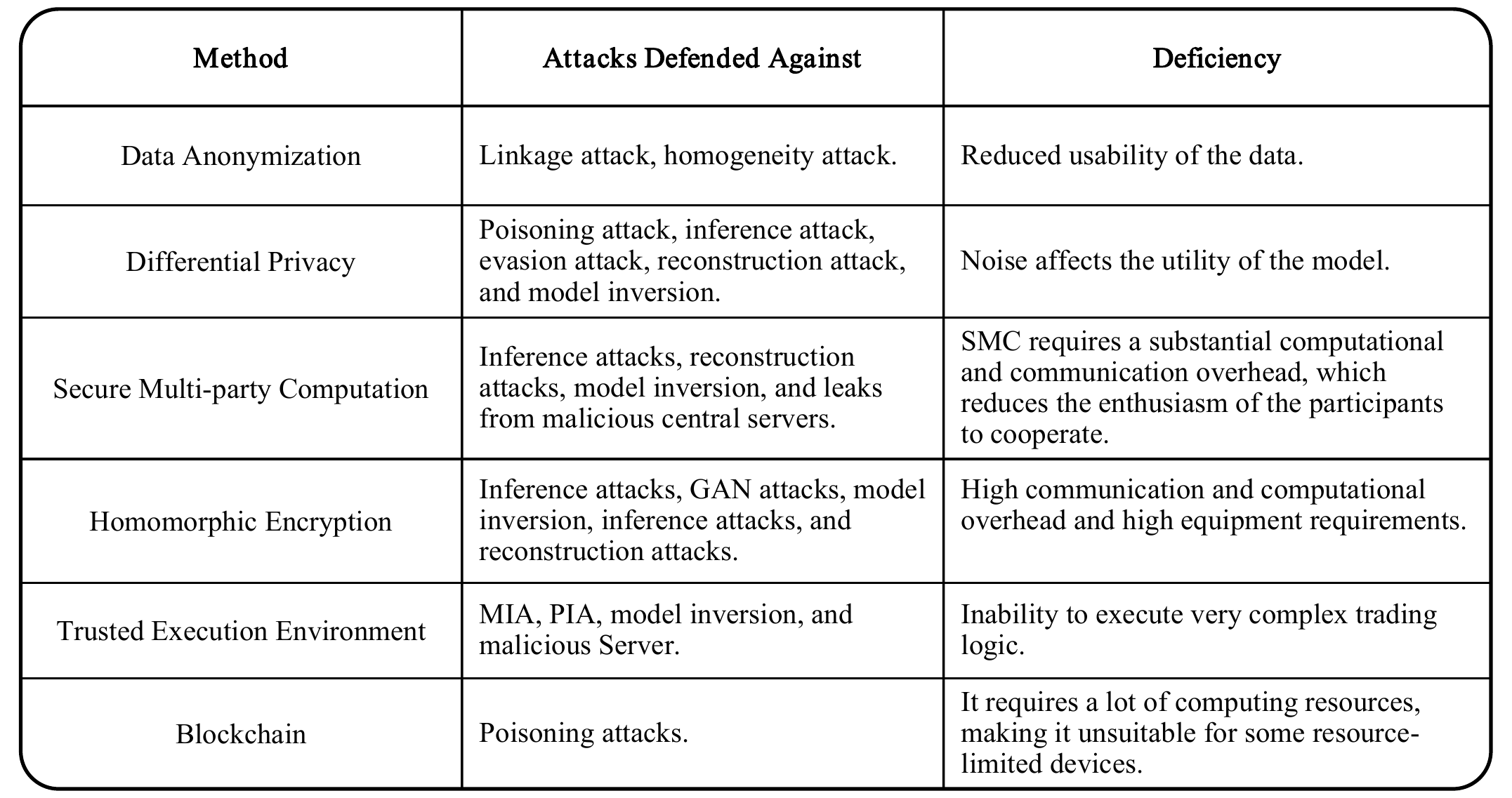}
    \label{fig:Defend}
\end{table}

\textbf{Data anonymization.} In order to defend against privacy attacks, we can use anonymization techniques to hide or remove sensitive personal attributes, such as personally identifiable information (PII) \cite{schwartz2011pii}, so that the attacker cannot identify a specific individual through the data. K-anonymity, L-diversity, and T-closeness are three common anonymization techniques. Anonymization techniques have been used to improve the privacy of FL \cite{song2020analyzing, choudhury2020syntactic}. This type of approach improves privacy by hiding or removing sensitive information, but may reduce the usability of the dataset \cite{merener2012theoretical}. In addition, much anonymized data can be easily ``de-anonymized". Thus, data anonymization is often used in conjunction with other ways to protect privacy.

\textbf{Differential privacy (DP)} \cite{dwork2008differential}. This is a common approach for protecting privacy in FL frameworks by adding noise to the uploaded data, making it impossible for an attacker to access the original data or model. It hinders the reverse data retrieval from the attacker \cite{gosselin2022privacy}. DP defends against attacks during the training phase and the prediction phase of FL. Those types of attacks that DP defends against are poisoning attack, inference attack, evasion attack, reconstruction attack, and model inversion \cite{truong2021privacy, bouacida2021vulnerabilities, zhang2022security}. Although DP can improve the privacy of the FL framework, the utility of the models can be seriously affected if too much noise is added. Therefore, how to balance privacy and utility is an important issue when using DP in the FL framework.
    
\textbf{Secure multi-party computation (SMC)} \cite{goldreich1998secure}. SMC is a generic cryptographic primitive for solving privacy-preserving collaborative computation problems between a set of mutually distrusting participants \cite{zhao2019secure}. SMC does not leak the input and output of the participant to other members participating in the computation. In the FL framework, SMC is able to defend against inference attacks, reconstruction attacks, model inversion, and leaks from malicious center servers \cite{zhang2021survey, zhang2022security}. SMC contains complex cryptographic operations, which leads to a large computational overhead and high performance loss. This may reduce the participants' interest in cooperating, so SMC is not suitable for large-scale FL scenarios.
    
\textbf{Homomorphic encryption (HE) \cite{paillier1999public}.} The principle that HE can protect data privacy is that it does not touch the original data. HE first encrypts the data, then processes it, and finally decrypts it. HE defends against attacks during the training phase and the prediction phase of FL. Although HE provides strict privacy guarantees, HE incurs a large computational overhead \cite{truong2021privacy}. Therefore, this approach is not suitable for FL scenarios with numerous participants and devices with limited computational resources.
    
\textbf{Trusted execution environment (TEE) \cite{sabt2015trusted}.} TEE is a tamper-proof and trusted ecosystem for executing authenticated and verified code. In the FL framework, TEE establishes digital trust by protecting devices in FL, which effectively prevents attackers from attacking private information \cite{bouacida2021vulnerabilities}. TEE can defend against attacks such as MIA, PIA, mode inversion, and malicious server \cite{zhang2022security}. However, TEE has limited execution space, which prevents complex transaction logic from being executed.
    
\textbf{Blockchain} \cite{zheng2017overview}. It is a distributed ledger technology that uses the blockchain to verify and store data, generates and updates data through a consensus mechanism, and involves an intelligent contract and an incentive mechanism \cite{zhang2021survey}. What's more, blockchain technology is decentralized, tamper-proof, unforgeable, auditable, and accountable \cite{issa2022blockchain}. Blockchain is a preferred solution to the problem of implementing data security and data validation in a non-centralized FL scenario \cite{zhang2021survey}. In addition, the verifiability of blockchain will reduce the impact of poisoning attacks \cite{gosselin2022privacy}. This method has the drawback that it can't be used in FL situations with a lot of people and devices with limited computing power.

\subsection{Defenses at the Security Level}

Vulnerabilities in FL provide an opportunity for curious attackers or malicious attackers to gain unauthorized access. To ensure the security of the FL framework, we want to scan for all vulnerabilities as much as possible. There are three main sources of vulnerabilities: insecure communication channels, malicious clients, and central parameter servers that are not robust or secure enough. The defense methods for security breaches can be divided into active and passive defenses. The purpose of active defense is to detect and mitigate the risk to the FL framework in advance, before it has an impact on the framework. The purpose of passive defense is to remediate and mitigate the impact when an attack has already occurred. The security defense approach is closely related to the CIA \cite{alazab2021federated}, which is the core of information security and lays the foundation for all security-based frameworks. The CIA refers to the three qualities of information security—confidentiality, integrity, and availability—that are usually used to analyze security defense methods in the FL framework.

\textbf{Confidentiality}. The types of attacks that compromise confidentiality are inference attack, reconstruction attack, and model extraction attack. Nasr \textit{et al.} \cite{nasr2019comprehensive} showed that data can be inferred by considering model weights in the FL environment. It has also been demonstrated that local datasets can be reconstructed by gradients \cite{geiping2020inverting}. In addition to the privacy-preserving approaches, many defensive approaches have been proposed to ensure the confidentiality of data. For example, VerifyNet, a verifiable FL framework, can guarantee the confidentiality of the gradients uploaded by participants using the proposed double masking protocols \cite{xu2019verifynet}.

\textbf{Integrality}. Poisoning attack and Evasion attack are types of attacks that compromise integrity. The primary purpose of ensuring data integrity is to ensure that once data is collected, it cannot be tampered with. Common methods to ensure data integrity are TEE and blockchain. TEE enables end-to-end security and authentication. People have applied TEE to the FL framework to detect participants who violate training integrity protocols \cite{chen2020training}. Blockchain is tamper-evident, decentralized, and protected against single points of failure, and these properties meet the requirements of integrity protection. In addition, there are several methods to ensure integrity by screening malicious clients \cite{rodriguez2020dynamic}.

\textbf{Availability}. The attacks on availability in the FL framework are related to Byzantine attacks. Until now, many kinds of defense methods against Byzantine attacks have been proposed. For example, the Krum algorithm is an aggregation rule with resilience properties that can be used to defend against Byzantine attacks \cite{blanchard2017machine}. Implementing incentives in the FL framework is a good way to improve data availability by rewarding or penalizing participants based on the value of their contributions. This reduces the possibility of participants sending useless or harmful data, and also improves the usability of the training model.

\textbf{Discussion:} It is clear from the above analysis that each defense method focuses on addressing one or more types of attack methods. There is no single defense method that can address all types of attacks. That is, if we want to keep the FL framework as secure and private as possible, we need to combine multiple defense methods into the FL framework in a harmonious way. How to choose the right defense method is an important issue. In addition to considering the defense capability, the utility of the data or model and cost are also important considerations. The defense method should not sacrifice the utility of the data or model; otherwise, it has little application value. In addition, defense methods require expensive equipment or cost, which greatly limits the application scenarios.

\section{Challenges and Promising Directions} 
\label{sec:challenges}

From the various attacks and defenses mentioned above, it can be seen that although the FL framework and the corresponding techniques can protect data to a certain extent, there are still many security issues to be solved. After sorting through and analyzing the relevant work from the past few years, we've come up with a list of several ways that could be improved and deserve more research:

\subsection{Trade-off on Security, Communication, and Computing}

While measures such as homomorphic encryption and differential privacy protect the security of privacy, they can also reduce FL performance, including increased communication delay and increased computing load. In future research, applying cloud computing to FL may be a way to optimize performance with the help of edge computing technology. In addition, recent related research considers choosing an appropriate security strategy \cite{hao2019towards, choquette2021capc}, with the expectation that it will maximize communication efficiency and computational efficiency as much as possible while ensuring system security.

\subsection{Emphasis on Robustness or Emphasis on Privacy?}

Current research on robustness and privacy is inherently conflicted \cite{lyu2020privacy}. On the one hand, since robustness pursues the universality of defenses against adversarial attacks, it is necessary to find commonalities between different attacks, which requires greater access to data and models. However, FL does not comply with the principles in GDPR\footnote{\url{https://en.wikipedia.org/wiki/General_Data_Protection_Regulation}}, a regulation designed to protect data, and the pursuit of robustness may exacerbate this problem \cite{truong2021privacy}. On the other hand, privacy pursues a comprehensive defense against a certain type of attack with encryption and other means, but may overlook opportunities for other types of attacks (e.g., a lack of robustness). Exploring how to strike a balance between them is a direction worthy of future research.

\subsection{Contribution Measurements and Incentive Strategies}

We know that it is not easy to guarantee that all clients are honest and well-intentioned. In the FL system, we define individuals who benefit from collective resources or public goods but do not pay anything as free-riders. In general, each client receives an incentive or reward for making a contribution within the FL system. In this way, those free-riders also get incentives for free by sending fake updates to the global model. Previous studies \cite{tseng2011free} extensively discussed and analyzed this phenomenon in peer-to-peer (P2P) networks. A defensive method called Standard deviation-Deep Autoencoding Gaussian Mixture Model (STD-DAGMM) \cite{lin2019free} was proposed to identify free-riders and prevent them from receiving updated models and rewards. As more and more workers emerge into the FL system, certain defensive strategies need to be adopted to deal with this malicious behavior for the purpose of improving the accuracy and fairness of the FL system. Therefore, it is necessary to explore other incentive strategies to detect fake model updates and assess the actual contribution of each client.

\subsection{Attacks and Defenses in the New Form of FL}

As mentioned above, the current research on attacks and defenses of FL mainly focuses on the most basic form, e.g., HFL with a homogeneous model architecture. For FL with heterogeneous model architecture \cite{litany2022federated, qu2022rethinking}, VFL, and FTL, the applicability of existing privacy-preserving techniques and attack defense mechanisms has not been studied fully and empirically. In addition, based on the innovative idea that each participant can become a server, decentralized FL has received relevant research \cite{korkmaz2020chain, hu2020gfl, li2021blockchain}. Does the rotation of server roles and permissions aggravate or distract from security risks? Security research in this area is also a future research direction.

\subsection{Another Interpretation of the Means of Attacks}

Note that attacks and defenses are inseparable. Only if the attack keeps developing can the defense keep improving. Thereafter, a thorough study of attacks can promote the development of defenses to a certain extent, thus promoting the development of FL. Apart from threatening models, some attack methods can also be used to study good technical applications. For example, Adi \textit{et al.} \cite{adi2018turning} found that the production, embedding, and verification in the backdoor attack were consistent with the three processes of the watermarking mechanism. Therefore, backdoor technology is used to watermark deep neural networks in a black-box way for intellectual property protection. With the in-depth study of attack and defense problems in FL, a promising research direction is how to use these technologies to address some of the limits of privacy protection in real life, in addition to focusing on attack and defense technology itself.

\subsection{Applications of the FL Systems}

It is key to strengthen the defensive strategy in the FL system, integrate privacy protection measures such as secure multi-party computing or differential privacy, and build an enhanced privacy protection FL system framework to simulate the attack and the defense in reality. There are a few open-source systems that provide FL frameworks for researchers and developers to continuously improve and upgrade them, including Google's TFF\footnote{\url{https://www.tensorflow.org/federated/federated_learning}}, PySyft\footnote{\url{https://blog.openmined.org/tag/pysyft/}} based on the PyTorch framework, and FATE\footnote{\url{https://fate.fedai.org/}} developed by WeBank. However, it should be noted that, besides PySyft, there is currently no framework or system that can incorporate and execute differential privacy or secure multi-party computation in real-world applications \cite{mothukuri2021survey}. Thereafter, integrating constantly updated privacy protection technologies into FL systems is a challenging direction. It will also be important to use FL systems in more areas of life.

\section{Conclusion}  \label{sec:conclusion}

Federated learning, as a new breed of artificial intelligence, is currently in a baby state. It follows local computing and model transmission, two concepts that reduce the cost and risk of privacy leakage brought by traditional centralized machine learning approaches. Although FL can solve some real-world problems, there are still many potential threats. Aiming at providing a comprehensive survey and giving readers a clear view and understanding of attacks and defenses in FL, we introduce and describe the existing work and research of the FL framework in five parts: background, characteristics, classification, systematic attack approaches, and systematic defense strategies. Then, we classify the possible attacks and threats according to the current three classification criteria, list the attack methods for each category, and introduce the corresponding principle of the specific attack. Later on, against these attacks and threats, the specific defense measures are summarized in two parts: privacy and security, respectively. Finally, we discuss six challenges in FL from the perspective of attacks and defenses. We also highlight several promising directions for future work in this quite active research area.

\section*{Acknowledgment}   

This research was supported in part by Key-Area Research and Development Program of Guangdong Province (No. 2020B0101090004), National Natural Science Foundation of China (Nos. 62002136, 62272196, 61932011), Natural Science Foundation of Guangdong Province (No. 2022A1515011861), Guangzhou Basic and Applied Basic Research Foundation (Nos. 202102020277, 2019B1515120010), Guangdong Key R\&D Plan2020 (No. 2020B0101090002), Fundamental Research Funds for the Central Universities of Jinan University (No. 21622416), the Young Scholar Program of Pazhou Lab (No. PZL2021KF0023), National Key Research and Development Plan of China (No. 2020YFB1005600), and National Joint Engineering Research Center for Network Security Detection and Protection Technology, Guangdong Key Laboratory for Data Security and Privacy Preserving. Dr. Wensheng Gan is the corresponding author of this paper.


\bibliographystyle{IEEEtran}
\bibliography{references}

\end{document}